\documentclass[10pt,twoside,notitlepage,a4paper]{article}
\usepackage{citesort}
\usepackage[bf, footnotesize]{caption}
\usepackage{graphicx}

\begin{document}
\title{Flat Histogram Method of Wang-Landau and N-Fold Way}
\author{B.J. Schulz, K. Binder and M. M\"uller \\ {\footnotesize Institut f\"ur Physik, WA331, Johannes Gutenberg Universit\"at} \\ {\footnotesize Staudinger Weg 7, D55099 Mainz, Germany}
}
\maketitle
\begin{abstract} We present a method for estimating the density of states of a classical statistical model. The algorithm successfully combines the Wang-Landau flat histogram method with the N-fold way in order to improve efficiency of the original single spin flip version. We test our implementation of the Wang-Landau method with the two-dimensional nearest neighbor Ising model for which we determine the tunneling time and the density of states on lattices with sizes up to $50\times 50$. Furthermore, we show that our new algorithm performs correctly at right edges of an energy interval over which the density of states is computed. This removes a disadvantage of the original single spin flip Wang-Landau method where results showed systematically higher errors in the density of states at right boundaries. In order to demonstrate the improvements made, we compare our data with the detailed numerical tests presented in a study by Wang and Swendsen where the original Wang-Landau method was tested against various other methods, especially the transition matrix Monte Carlo method (TMMC). Finally, we apply our method to a thin Ising film of size $32\times 32\times 6$ with antisymmetric surface fields. With the density of states obtained from the simulations we calculate canonical averages related to the energy such as internal energy, Gibbs free energy and entropy, but we also sample microcanonical averages during simulations in order to determine canonical averages of the susceptibility, the order parameter and its fourth order cumulant. We compare our results with simulational data obtained from a conventional Monte Carlo algorithm.  
\end{abstract}
\section[inro]{Introduction} 
While Monte Carlo methods in statistical thermodynamics already find broad application \cite{binder,frenkel,landau}, the standard approach using the Metropolis algorithm \cite{metro} has the important disadvantage that the entropy of the simulated model system is not an output of the calculation. Furthermore in systems with a rugged landscape of the (coarse-grained) free energy the convergence of the method in practice often is problematic - the system may get ``stuck'' in one valley of this free energy landscape for a long time.\\
In order to overcome these difficulties, many interesting approaches (e.g., umbrella sampling \cite{torrie}, multicanonical Monte Carlo \cite{berg}, expanded ensemble methods \cite{lyu,iba}) have been proposed in the literature, and some of these techniques emphasize the idea of directly sampling the energy density of states (e.g., \cite{lee,oliv,jswang1,wanglandau,jswang2,wangswendsen}). Such approaches are clearly very promising, and in particular the method of Wang and Landau \cite{wanglandau} has the additional merit that it is straightforward to implement. However, it is a difficult matter to judge the efficiency of the various methods, and estimate in beforehand the computational effort that is necessary to reach a desired level of accuracy for a chosen linear dimension $L$ of the (lattice) model under study. In particular, a recent comparative study of Wang and Swendsen \cite{wangswendsen} cautioned against a too optimistic view on these matters, and suggested that the Wang-Landau method \cite{wanglandau} for medium and large lattice sizes $L$ is considerably less efficient than various versions of transition matrix Monte Carlo methods \cite{wangswendsen}, which are more complicated to implement however.\\
In the present paper, we reconsider the Wang-Landau algorithm \cite{wanglandau} and modify it by combining it with the N-fold way algorithm of Bortz et al. \cite{bortz}. It has been known that this latter algorithm is far superior to the single spin flip Metropolis algorithm \cite{binder} at low temperatures. However, as will be demonstrated here, our new algorithm performs significantly better near the right edge of an energy interval, since for large $L$ it is necessary to split the total energy range in many subintervals (which are slightly overlapping, of course, so that they can be joined unambigously), which then can be sampled in parallel on the processors of a multi-processor machine. This improvement of accuracy on the right boundaries of energy intervals is crucial for obtaining a very good overall accuracy. Thus, we are able to show that this particular criticism does no longer apply with respect to this new algorithm and a performance of the algorithm comparable in efficiency to the transition matrix Monte Carlo methods is reached.\\
In sec. \ref{sec2}, the flat histogram method of Wang and Landau \cite{wanglandau} is briefly reviewed, as well as the N-fold way algorithm \cite{bortz}, and our combination of these two methods is presented. Sec. \ref{sec3} then describes the implementation and results for the two-dimensional Ising model, treated also in refs. \cite{wanglandau,jswang2,wangswendsen}, and a comparative analysis of errors in the density of states is presented. Sec. \ref{sec4} then gives, as an example of an application to a nontrivial model of current interest \cite{binderlandau,binderevans,ferrenberg} results for a three-dimensional Ising model in a $L \times L \times D$ thin film geometry ($L=32$, $D=6$), where competing boundary fields ($h_1=-h_2$) act on the two free $L \times L$ surfaces. It is shown that the present method produces results compatible with those of the conventional single spin flip algorithm \cite{binderlandau} and reaches a better accuracy with less effort in computing time. Sec. \ref{sec5} finally summarizes some conclusions.

\section[flat]{The flat histogram method of Wang-Landau} \label{sec2}
\subsection[single]{Single spin flip} \label{sifli}
Recently, F. Wang and Landau \cite{wanglandau} proposed a Monte Carlo algorithm for
classical statistical models which uses a random walk in energy space in order 
to obtain an accurate estimate for the density of states $g(E)$. This method
is based upon the fact that a flat energy histogram $H(E)$ is produced if the probability for the transition to a state of energy $E$ is proportional to $1/g(E)$.\\
This observation is utilized in the following way. Initially, $g(E)$ is set
equal to one for all energies. A spin is then chosen at random and flipped
with probability $\min (1,g(E)/g(E'))$ whereby $E'$ is the energy of the
system with the chosen spin being overturned. The density of states $g(E)$ is
not constant during the random walk, but is updated according to
$g(E)\rightarrow g(E)\cdot f$ after each spin flip trial whether the spin is
flipped or not. A histogram H(E) records how often a state of energy
$E$ is visited. In the beginning of the random walk the modification factor $f$ can be as large as $e\simeq2.7182818$.\footnote{If one chooses to sample $S(E)=\log g(E)$ the modification factor becomes an increment $S(E)\rightarrow S(E)+\log f$ with $\log f \leq \log e \simeq 0.4342944$.}Each time the energy histogram satisfies a certain flatness criterion, $f$ is reduced according to $f\rightarrow f^{\frac{1}{2}}$ and $H(E)$ is set to zero for all energies.
The histogram is considered as flat if 
\begin{equation}
H(E)\geq\epsilon\cdot\langle H(E)\rangle
\label{flatness} 
\end{equation}
for all $E$ where $\epsilon$ is usually between $0.7$ and $0.95$ and $\langle
H(E)\rangle$ is the average histogram. The simulation is ended if $f$ is close
enough to one, i.e., smaller than some predetermined final modification factor
$f_{final}$. To speed up simulations it is possible to perform several random
walks on adjacent energy intervals on independent processors. Disadvantages of
the single spin flip version of this algorithm are the small acceptance rates
for energy intervals which contain the groundstate and low-lying excited
states and the relatively large errors of $g(E)$ at right edges of energy
intervals as reported in \cite{wanglandau}. Since this flat histogram method
produces only a relative density of states $g(E)$, one has to normalize
$g(E)$ in order to get the absolute density of states $\hat{g}(E)$. 
This can be done by using known values of the density of states, for
example, the groundstate degeneracy or other constraints on $\hat{g}(E)$. In case of
the two-dimensional nearest neighbor Ising model 
\begin{equation}
\mathcal{H}=-J\sum\limits_{\langle i,j\rangle}s_i s_j \quad \mbox{with}\quad s_i=\pm 1,
\label{2i}
\end{equation}
one has
\begin{eqnarray}
\hat{g}(-2JN) & = & 2, \label{c1} \\
\hat{g}(-E) & = & \hat{g}(E), \label{c2} \\ 
\sum\limits_{E} \hat{g}(E) & = & 2^{N}, \label{c3} 
 \end{eqnarray}
where $N$ is the number of spins. If one uses the groundstate to normalize $g(E)$ one can calculate canonical averages from the density of states for all temperatures which become exact for $T\rightarrow 0$. It is thus highly desirable to estimate the density of states for low-lying energies with sufficient accuracy. Since we deal with low acceptance rates of the single spin flip approach in these energy-ranges one can reduce the simulational effort enormously by using a rejection-free update scheme. In \cite{wangswendsen} J. S. Wang and Swendsen presented an efficient Monte Carlo method using the N-fold way to obtain an estimate for the transition matrix from which the density of states can be determined via an optimization procedure. They have tested their algorithm with the two-dimensional Ising model and compared it to various other methods from which the density of states can be calculated. In particular, they found their method to perform better than the single spin flip Wang-Landau flat histogram method. Adopting the idea of Wang and Landau in the context of transition matrix they improved the efficency of the original algorithm only for small system sizes up to $L=8$, but for larger system sizes they reported that one has the problem of sticking to a Gaussian distribution for the histogram.
In the next subsection we will discuss how the algorithm of Wang and Landau can be successfully combined with the rejection-free N-fold way in order to enhance its performance.
\subsection[nfold]{N-fold way}
In the N-fold way \cite{bortz} a flip occurs at each step of the algorithm and one then calculates the life-time of the resulting state. The usual single spin flip dynamics is thus preserved and observables become life-time weighted averages over the generated states. We will now describe the method in the context of the algorithm of Wang and Landau.
Since the density of states $\hat{g}(E)$ can be very large especially for large
system sizes we consider $S(E)=\log g(E)$ during simulations. In the beginning $S(E)$ is set to
zero for all $E$. Initially the system may be in the state $\sigma$ with
energy $E\in I=\left[E_{min},E_{max}\right]$, whereby $I$ denotes the
energy-range for which one wants to estimate $g(E)$. One then partitions all
spins into classes according to their energetic local environment, i.e., the
energy difference $\Delta E_{i}$ a spin flip will cause. For a two-dimensional
nearest neighbor Ising model each spin belongs to one of only $M=10$ classes. 
The total probability $P$ of any spin of class $i$ being overturned is given by
\begin{equation}
P(\Delta E_{i})=n(\sigma,\Delta E_{i})p(E\rightarrow E+\Delta E_{i}),\quad i=1,...,M,  
\label{totp}
\end{equation}
with $n(\sigma,\Delta E_{i})$ being the number of spins of state $\sigma$ which
belong to class $i$ and $p(E\rightarrow E+\Delta E_{i})$ is given by
\begin{equation}
p(E\rightarrow E+\Delta E_{i})=\left\{ \begin{array}{r@{\quad \mbox{if} \quad}l}
                               \min (1,g(E)/g(E+\Delta E_{i})) & E+\Delta E_{i}\in
                               I \\ 0 &
                               E+\Delta E_{i} \not\in I. \end{array} \right.
\label{flira}
\end{equation}
In order to determine the class from which to flip a spin one
calculates the numbers
\begin{equation}
Q_{m}=\sum \limits_{i\leq m}P(\Delta E_{i}),\quad m=1,...,M\quad \mbox{and} \quad Q_{0}=0,
\label{q}
\end{equation}
which are the integrated probabilities for a spin flip within the first $m$
classes. Hence a class is selected by generating a random number $0<r<Q_{M}$
and taking class $m$ if $\, Q_{m-1}<r<Q_{m}$. The spin to be overturned
is chosen from this class with equal probabilities. Due to the flip, the spin and its interacting neighbors will change classes and correspondingly the numbers
 $n(\sigma,\Delta E_{i})$ will differ from their predecessors. 
Finally, one has to
determine the average life-time $\tau$ of the resulting state, i.e., one has to
find out how many times the move just made would be rejected on average in the usual
update scheme. The probability that the first random number would produce a flip is $\hat{P}=Q_{M}/N$. Therefore one has for the probability that exactly $n$ random numbers will result in a new configuration  
\begin{equation}
\bar{P}_{n}=\hat{P}(1-\hat{P})^{n-1}.
\end{equation}
Thus the average life-time becomes
\begin{equation}
\tau=\sum\limits^{\infty}_{n=1} n\bar{P}_{n}=\sum\limits^{\infty}_{n=1} n\hat{P}(1-\hat{P})^{n-1}= \frac{N}{Q_{M}}.
\label{liti}
\end{equation}
Now, the steps which are actually carried out by the N-fold way version of the Wang-Landau method are the following:
\begin{enumerate}
\item choose an initial configuration and set $H(E)=0$, $S(E)=0$ for all $E$ and $f_0=\log e\simeq 0.4342944$ and also fix $f_{final}$.
\item determine (update) the probabilities $p(E\rightarrow E+\Delta E_{i})$ 
and the $Q_{m}$'s of the (initial) configuration using eqs. (\ref{totp}), (\ref{flira}) and (\ref{q}).
\item determine average lifetime $\tau$ of (initial) configuration via eq. (\ref{liti}).
\item increment histogram, density of states and update $f_{i}$:
\begin{eqnarray}
H(E) & \rightarrow & H(E)+\tau \nonumber \\
S(E) & \rightarrow & S(E)+\Delta S(E) \nonumber\\
f_{i} & \rightarrow & f_{i+1}  \nonumber
\end{eqnarray}
with: 
\begin{eqnarray}
\Delta S(E) & = & \left\{\begin{array}{r@{\quad \mbox{if} \quad}l}f_{i} \cdot \tau & f_{i} \cdot \tau \leq \log e\simeq 0.4342944 \\ \log e &  f_{i} \cdot \tau > \log e \end{array}\right. \label{ds}  \\
f_{i+1} & = & \left\{ \begin{array}{r@{\quad \mbox{if} \quad}l}f_{i} & f_{i} \cdot \tau \leq \log e \\ \Delta S(E)/\tau & f_{i} \cdot \tau > \log e, \end{array} \right. \label{f1}   
\end{eqnarray}
in case of $f_i/f_{i+1}>2$ we set $f_{i+1}\rightarrow f_i/2$.
\item after some fixed sweeps check $H(E)$ and refine $f_{j}$ according to $f_{j+1}=f_{j}/2$ if $H(E)$ is flat. Stop if histogram is flat for a $f_{i}\leq f_{final}$. 
\item determine the $Q_{m}$'s (the $p(E\rightarrow E+\Delta E_{i})$'s are not updated here).
\item choose and flip spin as described above.
\item go to $2.$
\end{enumerate}
Equations (\ref{ds}) and (\ref{f1}) ensure that the increment $\Delta S(E)$ is kept below or equal to $\log e$. In the single spin flip case this constraint is always guaranteed by the choice of the initial modification factor. Larger upper limits for $\Delta S(E)$ result in large statistical errors in $\hat{g}(E)$ as was reported in \cite{wanglandau}. Furthermore they can lead to $Q_{M}=0$ in the context of the N-fold way which will definitely destroy the iteration procedure described above. We have observed that adjusting $f_i$ according to eqs. (\ref{ds}) and (\ref{f1}) only applies to the first stage of iteration. Once the histogram is flat for the first time the increment $\Delta S(E)$ is simply $f_{i} \cdot \tau$. Controling $f_{i}$ in other ways may also work.
\section[sim]{Simulations} \label{sec3}
\subsection[2d]{Two-dimensional Ising model}
First we have tested our algorithm with the two-dimensional Ising model, eq.
(\ref{2i}), on $L\times L$ square lattices with $L=8,16,32,50$ over the entire
energy-range $E/JN\in [-2,2]$. For these system sizes, the density of states
$\hat{g}(E)$ can be calculated exactly \cite{beale} and we will denote it by $\hat{g}_{ex}(E)$. As already mentioned the method only provides a relative density of states. To obtain the absolute density of states $\hat{g}(E)$, we first utilize (\ref{c2}) by taking
\begin{equation}
g(E)\rightarrow (g(E)+g(-E))/2.  
\end{equation}
In order to normalize $g(E)$ one can now use eq. (\ref{c1}) which will yield the
best accuracy in the density of states for low and high-lying energy levels or
one uses eq. (\ref{c3}) which will lead to best accuracy in $\hat{g}(E)$ for energies for which
the corresponding density of states has large values.
We have to stress here that this is the simplest method to obtain an absolute density
of states, but it suffices in order to demonstrate the improvements made by
combining the Wang-Landau method with the N-fold way. For further improvement
of accuracy concerning the density of states, one has to resort to
optimization methods. This was done succesfully in \cite{wangswendsen}, where a least-squares
method was considered with eqs. (\ref{c1}),(\ref{c2}) and (\ref{c3}) as constraints. 
 
\begin{figure}[f]
\begin{center}
\includegraphics[scale=0.50,angle=-90]{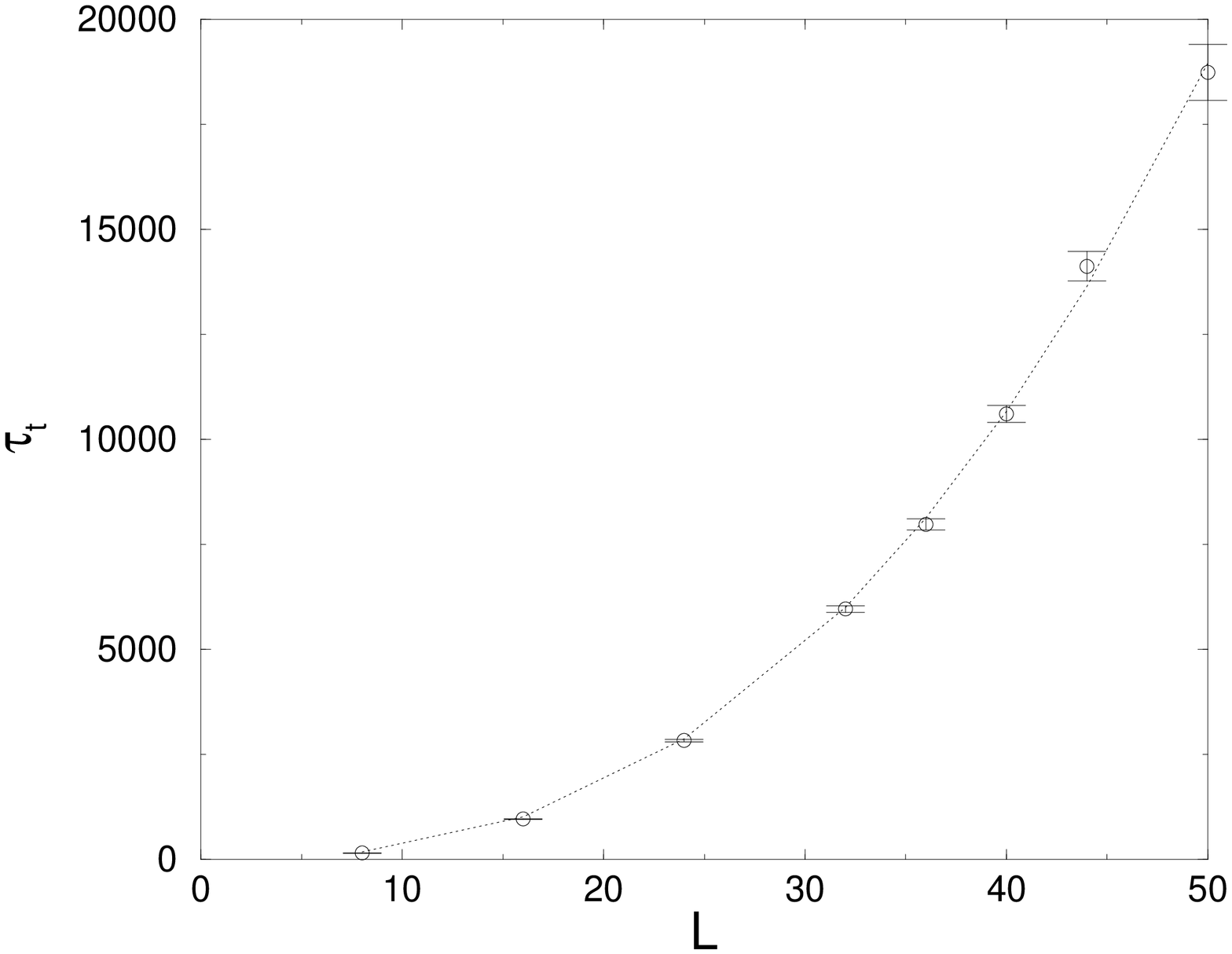}
\caption{Tunneling-time $\tau_{t}$ of the two-dimensional Ising model in units of sweeps. $\tau_t$ was fitted according to $x\cdot L^{\alpha}$ which yielded $x\simeq0.79$ and $\alpha \simeq 2.58$. We have used $\epsilon=0.85$ and $f_{final}\simeq 1.0 \cdot 10^{-6}$. The indicated error bars result from averages over 30 runs ($L=8,16,24,32,36$) and 10 runs ($L=40,44,50$).}
\label{tunn}
\end{center}
\end{figure}

\begin{figure}[f]
\begin{center}
\includegraphics[scale=0.50,angle=-90]{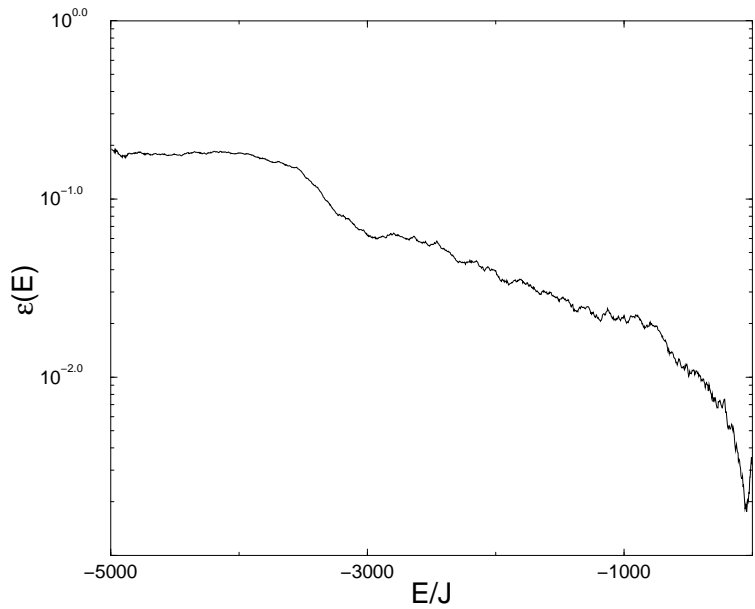}
\caption{Relative error $\varepsilon (E)$ in the density of states
  $\hat{g}(E)$ for the two-dimensional Ising model of size $L=50$. The error
  $\varepsilon (E)$ is an average over $30$ runs. The average number of sweeps
  was $(9.1\pm 0.4)\cdot 10^{5}$. We have used $\epsilon=0.7$ and
  $f_{final}=(7.6\pm 0.2)\cdot 10^{-7}$. The average relative error is
  $\bar{\varepsilon}=0.080$. The mirror image from $E/J=0$ up to $E/J=5000$ is not depicted.}
\label{err50}
\end{center}
\end{figure}

\begin{figure}[f]
\begin{center}
\includegraphics[scale=0.50,angle=-90]{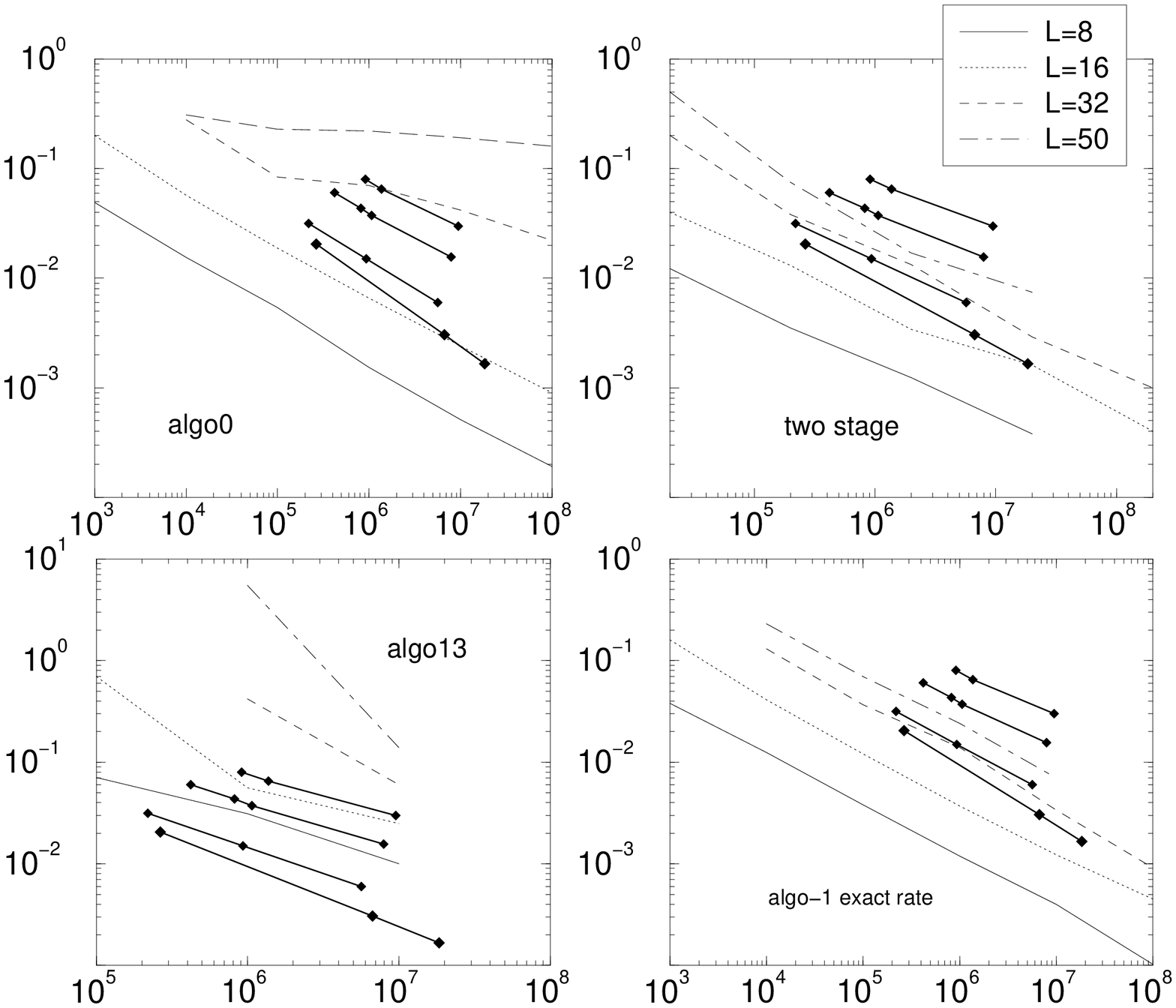}
\caption{Average error $\bar{\varepsilon}$ in the density of states $\hat{g}(E)$ over the number of sweeps, as listed in table \ref{table2di} (thick solid lines correspond to $L=50,32,16,8$ from top to bottom) in comparison with the data provided in \cite{wangswendsen} for the same system sizes (see legend on right top) produced by the algorithms\newline
{\bf algo0}: TMMC with original flat histogram flip rates.\newline
{\bf algo-1 exact rate}: Also a TMMC with exact multi-canonical flip rates.\newline
{\bf two stage}: Algorithm \emph{algo0} followed by \emph{algo-1} which uses 
the estimated density of states from the first stage for the multi-canonical 
flip rates.\newline
{\bf algo13}: Implementation of the Wang-Landau flat histogram method using 
single spin flips.}
\label{errcomp}
\end{center}
\end{figure}

To compare our data with the
tests provided in \cite{wangswendsen} we consider the relative error per energy level for the density of states 
\begin{equation}
\varepsilon(E)=\left|\frac{\hat{g}(E)}{\hat{g}_{ex}(E)}-1\right|
\end{equation}
and its average
\begin{equation}
\bar{\varepsilon}=\frac{1}{N-2}\sum\limits_{E}\varepsilon(E).
\end{equation}
The factor $1/(N-2)$ is due to the fact that we have $N-1$ different energy levels whereby the groundstate is excluded from averaging because we have imposed the exact value for the corresponding density of states, see eq. (\ref{c1}).
For each system size we have performed $30$ runs, starting always with all
spins up. Monte Carlo time was measured in units of sweeps, whereby one sweep was taken to be N spin flips. The results are given in table \ref{table2di} and figures \ref{err50} and \ref{errcomp}. The used parameters $\epsilon$ and $f_{final}$ are stated in the corresponding captions.
We also measured the tunneling-time $\tau_{t}$, i.e., the average number of 
sweeps which is elapsing during a transition from the groundstate to a state
of highest energy or vice versa. The tunneling time is averaged over the complete run, i.e., the various stages with different modification increments. For the two-dimensional Ising model 
we get
\begin{equation}
\tau_{t}\simeq 0.79L^{2.58}
\end{equation}
from a fit which is depicted in figure \ref{tunn}. For comparison,
the transition matrix Monte Carlo method yielded \cite{wangswendsen}
\begin{equation}
\tau_{t}\simeq 0.4L^{2.8}.
\end{equation}
We have to add here that the tunneling-time is difficult to define within the Wang-Landau method because
it depends on the density of states which is not constant during the simulation. Thus $\tau_t$ may also
depend on the interplay between the parameters $\epsilon$ and $f_{final}$. Thus we have chosen the typical values $\epsilon=0.85$ and $f_{final}\simeq 1.0\cdot 10^{-6}$ for all considered system sizes.

\begin{table}[f]
\begin{center}
\small
\begin{tabular}{|c|c|c|c|c|c|} \hline
 $L$ & sweeps & $\epsilon$ & $f_{final}$ & checks & $\bar{\varepsilon}$ \\ \hline\hline
  $8$  &$(2.652\pm 0.087)\cdot 10^{5}$&$0.97$&$(7.1\pm 0.3)\cdot 10^{-6}$&$1000$&$0.020$ \\ \hline
  $8$  &$(6.7\pm 0.8)\cdot 10^{6}$&$0.98$&$(7.0\pm 0.2)\cdot10^{-10}$&$1000$&$0.0030$ \\ \hline
  $8$  &$(1.840\pm 0.095)\cdot 10^{7}$&$0.99$&$(7.0\pm 0.2)\cdot10^{-10}$&$1000$&$0.0017$ \\ \hline\hline
  $16$  &$(2.19\pm 0.16)\cdot 10^{5}$&$0.8$&$(7.7\pm 0.3)\cdot10^{-7}$&$100$&$0.031$  \\ \hline
  $16$  &$(9.31\pm 0.42)\cdot 10^{5}$&$0.9$&$(7.0\pm 0.3)\cdot10^{-8}$&$100$&$0.015$ \\ \hline
  $16$  &$(5.7\pm 0.6)\cdot 10^{6}$&$0.95$&$(7.0\pm 0.3)\cdot10^{-9}$&$100$&$0.0060$ \\ \hline\hline
  $32$  &$(4.2\pm 0.2)\cdot 10^{5}$&$0.7$&$(7.1\pm 0.3)\cdot10^{-7}$&$100$&$0.060$  \\ \hline
  $32$  &$(8.19\pm 0.46)\cdot 10^{5}$&$0.8$&$(7.4\pm 0.3)\cdot10^{-7}$&$100$&$0.042$ \\  \hline
  $32$  &$(1.065\pm 0.042)\cdot 10^{6}$&$0.85$&$(7.3\pm 0.3)\cdot10^{-7}$&$100$&$0.037$ \\ \hline
  $32$  &$(7.9\pm 0.6)\cdot 10^{6}$&$0.9$&$(6.7\pm 0.3)\cdot10^{-9}$&$100$&$0.016$ \\ \hline\hline
  $\textbf{50}$&$(9.1\pm 0.4)\cdot 10^{5}$&$0.7$&$(7.6\pm 0.2)\cdot10^{-7}$&$100$&$0.080$  \\  \hline
  $50$  &$(1.371\pm 0.067)\cdot 10^{6}$&$0.85$&$(7.4\pm 0.2)\cdot10^{-7}$&$100$&$0.065$      \\ \hline
  $50$  &$(9.5\pm 0.8)\cdot 10^{6}$&$0.9$&$(5.9\pm 0.5)\cdot10^{-8}$&$100$&$0.030$   \\ \hline 
\end{tabular}
\caption{Simulation data for the two-dimensional Ising model. The number of sweeps, $f_{final}$ 
and the relative error $\bar{\varepsilon}$ of the density of states $\hat{g}(E)$ are averages over $30$ runs. 
Under \emph{checks} we have listed the number of sweeps between two successive checks of the 
histogram $H(E)$ for flatness. $\hat{g}(E)$ was obtained by normalizing with respect to the total number of states, eq. (\ref{c3}). $\varepsilon(E)$ for $L=50$ (bold) is depicted in figure \ref{err50}. 
See figure \ref{errcomp} for comparison with the data provided in \cite{wangswendsen}. }
\label{table2di}
\end{center}
\end{table}

\subsection[errright]{Errors at right edges}
\begin{figure}[f]
\begin{center}
\includegraphics[scale=0.50,angle=-90]{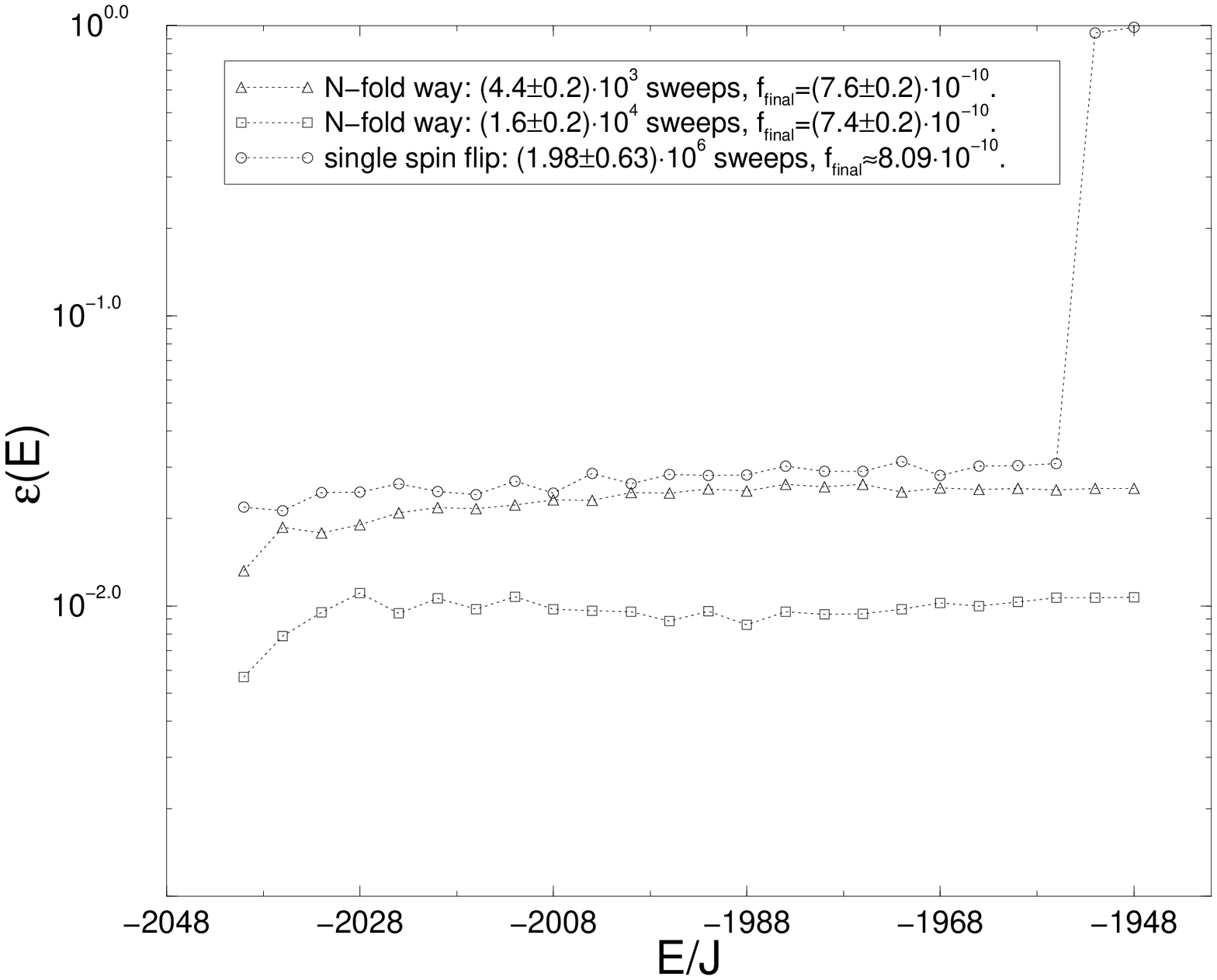}
\caption{Relative error $\varepsilon (E)$ in the density of states $\hat{g}(E)$ of the first $25$ energy levels of a two-dimensional Ising model with $L=32$. $\hat{g}(E)$ was obtained by normalizing with respect to the groundstate, $\varepsilon (E)$ is an average over $30$ runs. We have used $\epsilon=0.95$ (circles and triangles) and $\epsilon=0.98$ (squares). The N-fold way was approximately $35$ times faster (triangles) and $10$ times faster (squares) concerning CPU time than the single spin flip approach.}
\label{edges}
\end{center}
\end{figure}

As already mentioned in section \ref{sifli}, one shortcoming of the single spin
flip Wang-Landau method are the relatively large errors in the density of
states at right edges of an energy interval over which one wants to determine 
the density of states \cite{wanglandau}. In the N-fold way version presented here such systematic errors do not occur. In
order to demonstrate this we have calculated the density of states for the
first $25$ levels of a $L=32$ two-dimensional Ising model using single spin
flips as well as N-fold way updates, see figure \ref{edges}. This test also shows that the
simulational effort is enormously reduced in regions where the acceptance rate
for a spin flip is low.\\
In the single spin flip Wang-Landau method systematic errors occur at right edges due to a boundary effect which sets in as soon as the energy interval does not cover all the possible energies the system may have. If the random walk then hits an energy level which is outside the allowed range this move is simply rejected and the old level is counted once more analogous to the case when no boundary is involved in a transition, i.e., no difference is made between the density of states at boundaries and those away from the boundary, whereas in our N-fold way algorithm the density of states at edges is treated correctly since it is forbidden by the definition of the flip rates (eq. (\ref{flira})) to choose a spin whose flipping would result in an energy outside the allowed range.

\section[isfi]{Thin Ising films} \label{sec4}
\begin{figure}[f]
\begin{center}
\includegraphics[scale=0.50]{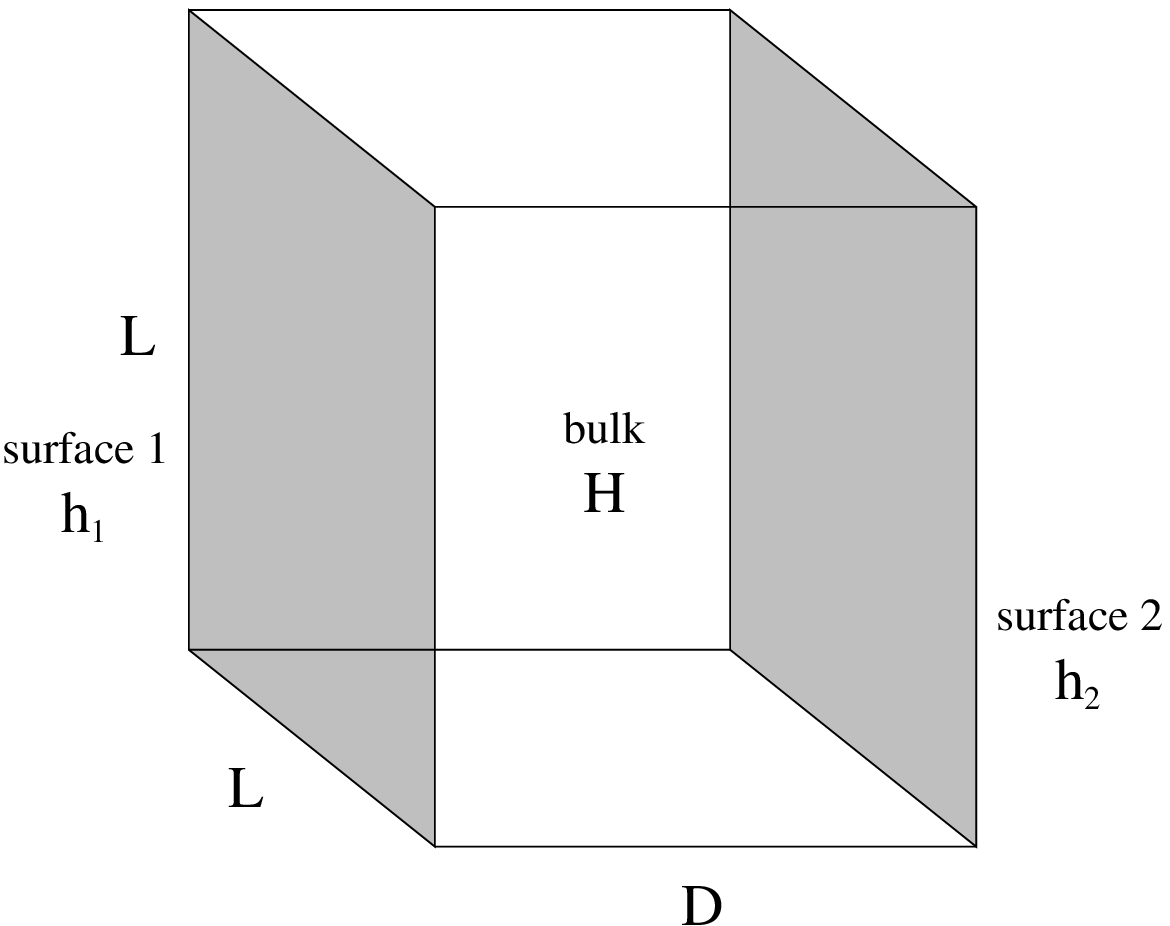}
\caption{Thin film geometry with magnetic fields $h_1$, $h_2$ acting on the surfaces (shaded) and a field $H$ acting on the bulk. Parallel to the surfaces periodic boundary conditions are imposed.}
\label{thingeo}
\end{center}
\end{figure}

\begin{figure}[f]
\begin{center}
\includegraphics[scale=0.50,angle=-90]{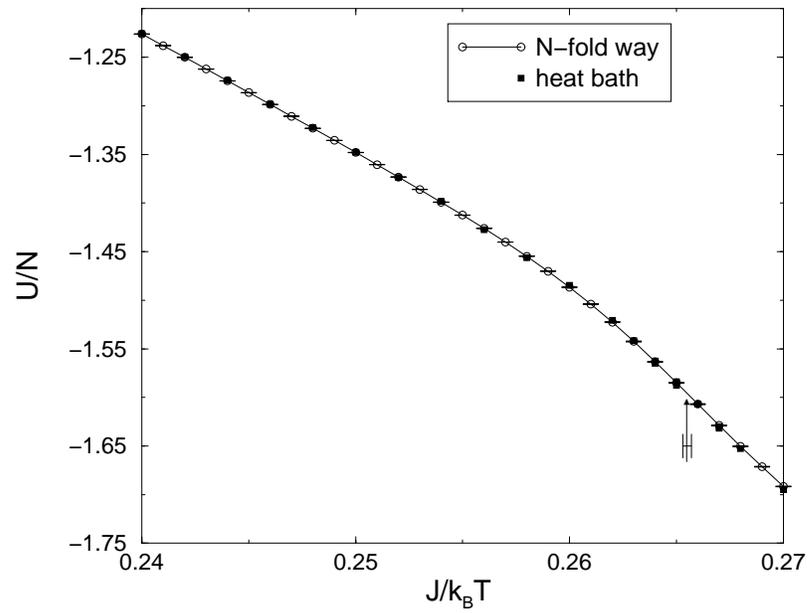}
\caption{Internal energy $U/N$ of a thin Ising film (eq. (\ref{thini})) 
with $L=32$ and $D=6$. The indicated error bars resulted from an average over 
$5$ runs. The average relative error of $U/N$ in the depicted range of $J/k_{B}T$ is $0.017\%$. Lines are only guides to the eye. The squares are data taken from ref. \cite{binderlandau}. The arrow marks the critical temperature \cite{binderlandau}. }
\label{intE}
\end{center}
\end{figure}

\begin{figure}[f]
\begin{center}
\includegraphics[scale=0.50,angle=-90]{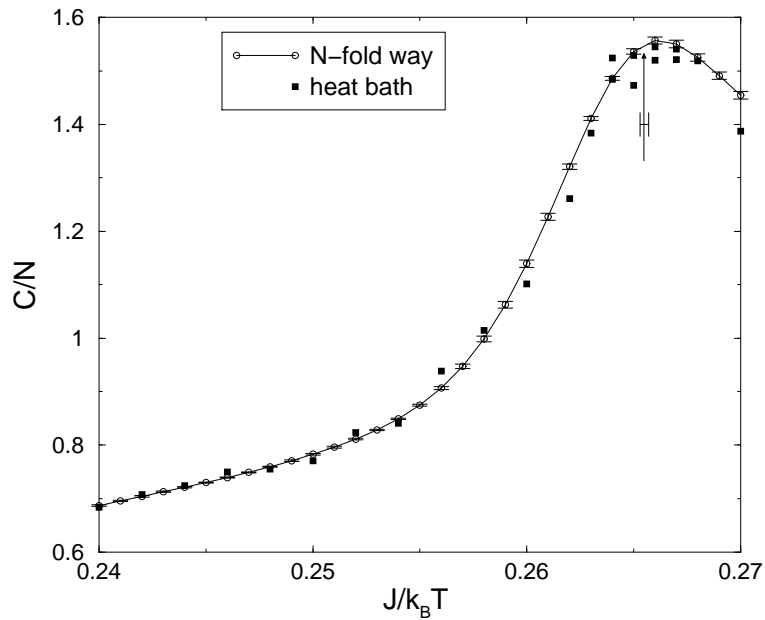}
\caption{Specific Heat $C/N$ of a thin Ising film with $L=32$ and $D=6$. The indicated error bars resulted from an average over $5$ runs. The average relative error of $C/N$ in the depicted range of $J/k_{B}T$ is $0.3\%$. The squares are data taken from ref. \cite{binderlandau}.}
\label{specH}
\end{center}
\end{figure}

\begin{figure}[f]
\begin{center}
\includegraphics[scale=0.50,angle=-90]{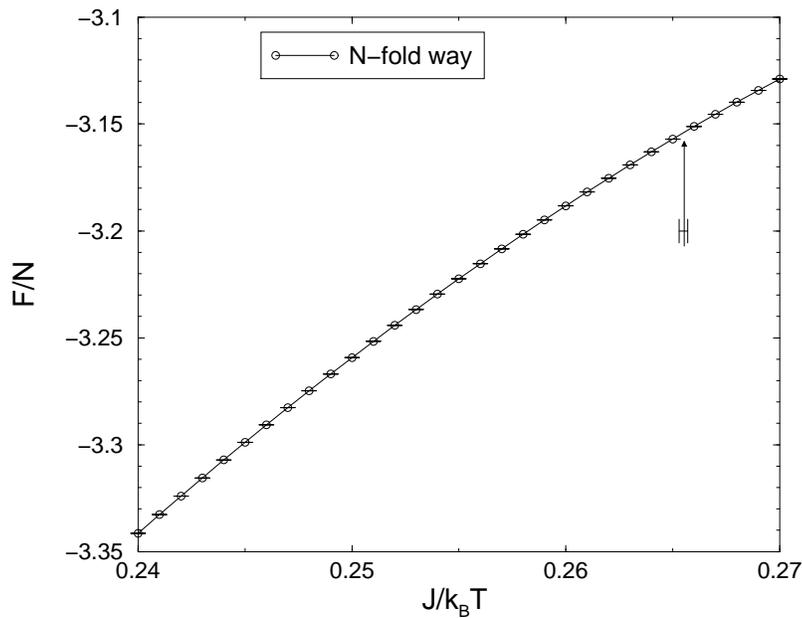}
\caption{Gibbs free energy $F/N$ of a thin Ising film with $L=32$ and $D=6$. The indicated error bars resulted from an average over $5$ runs. The average relative error of $C/N$ in the depicted range of $J/k_{B}T$ is $0.0011\%$.}
\label{gibbsF}
\end{center}
\end{figure}

\begin{figure}[f]
\begin{center}
\includegraphics[scale=0.50,angle=-90]{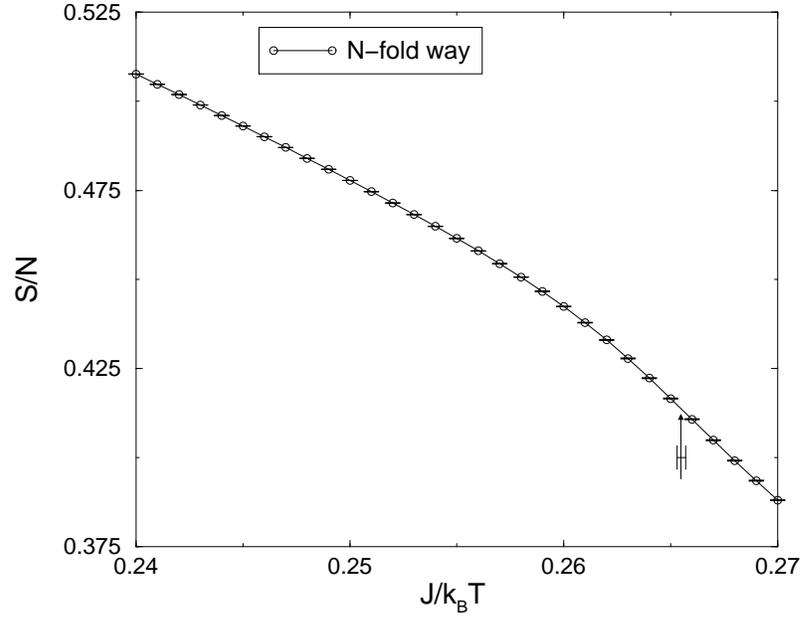}
\caption{Entropy $S/N$ of a thin Ising film with $L=32$ and $D=6$. The indicated error bars resulted from an average over $5$ runs. The average relative error of $S/N$ in the depicted range of $J/k_{B}T$ is $0.015\%$.}
\label{entropy}
\end{center}
\end{figure}

\begin{figure}[f]
\begin{center}
\includegraphics[scale=0.50,angle=-90]{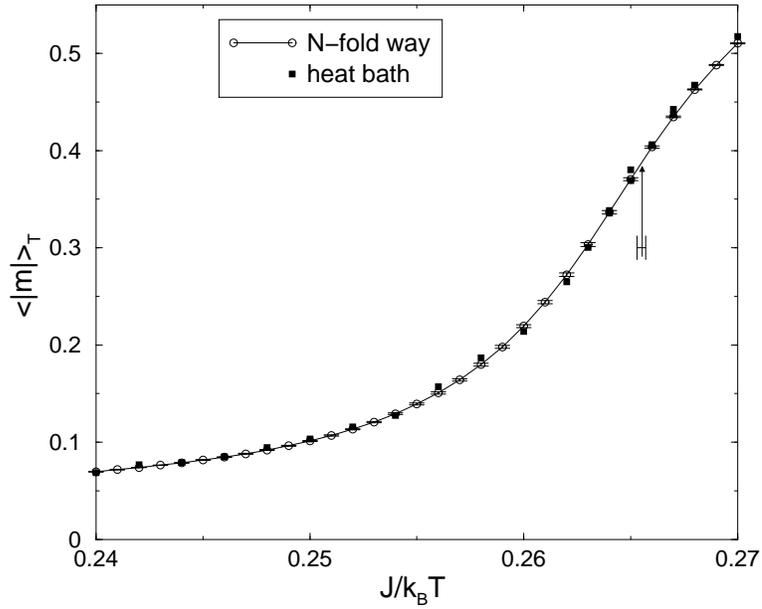}
\caption{Order parameter $\langle \left|m \right| \rangle_{T}$ of a thin Ising film with $L=32$ and $D=6$. The indicated error bars resulted from an average over $5$ runs.  The average relative error of $\langle \left|m \right| \rangle_{T}$ in the depicted range of $J/k_{B}T$ is $0.5\%$. The squares are data taken from ref. \cite{binderlandau}.}
\label{orderpar}
\end{center}
\end{figure}

\begin{figure}[f]
\begin{center}
\includegraphics[scale=0.50,angle=-90]{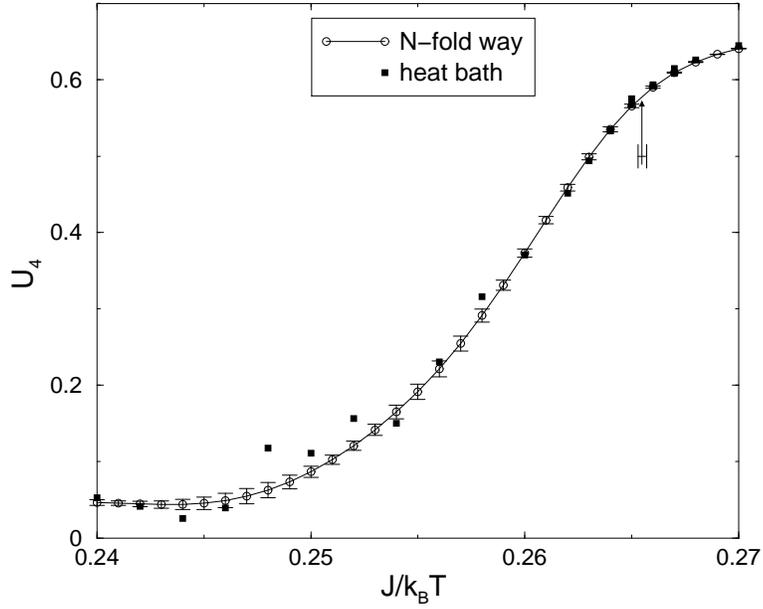}
\caption{Fourth order cumulant $U_{4}$  of the order parameter of a thin Ising film with $L=32$ and $D=6$. The indicated error bars resulted from an average over $5$ runs. The average relative error of $U_{4}$ in the depicted range of $J/k_{B}T$ is $6\%$. For $0.26\leq J/k_{B}T \leq 0.27$ it drops to $0.5\%$. The squares are data taken from ref. \cite{binderlandau}.}
\label{cumu}
\end{center}
\end{figure}

\begin{figure}[f]
\begin{center}
\includegraphics[scale=0.50,angle=-90]{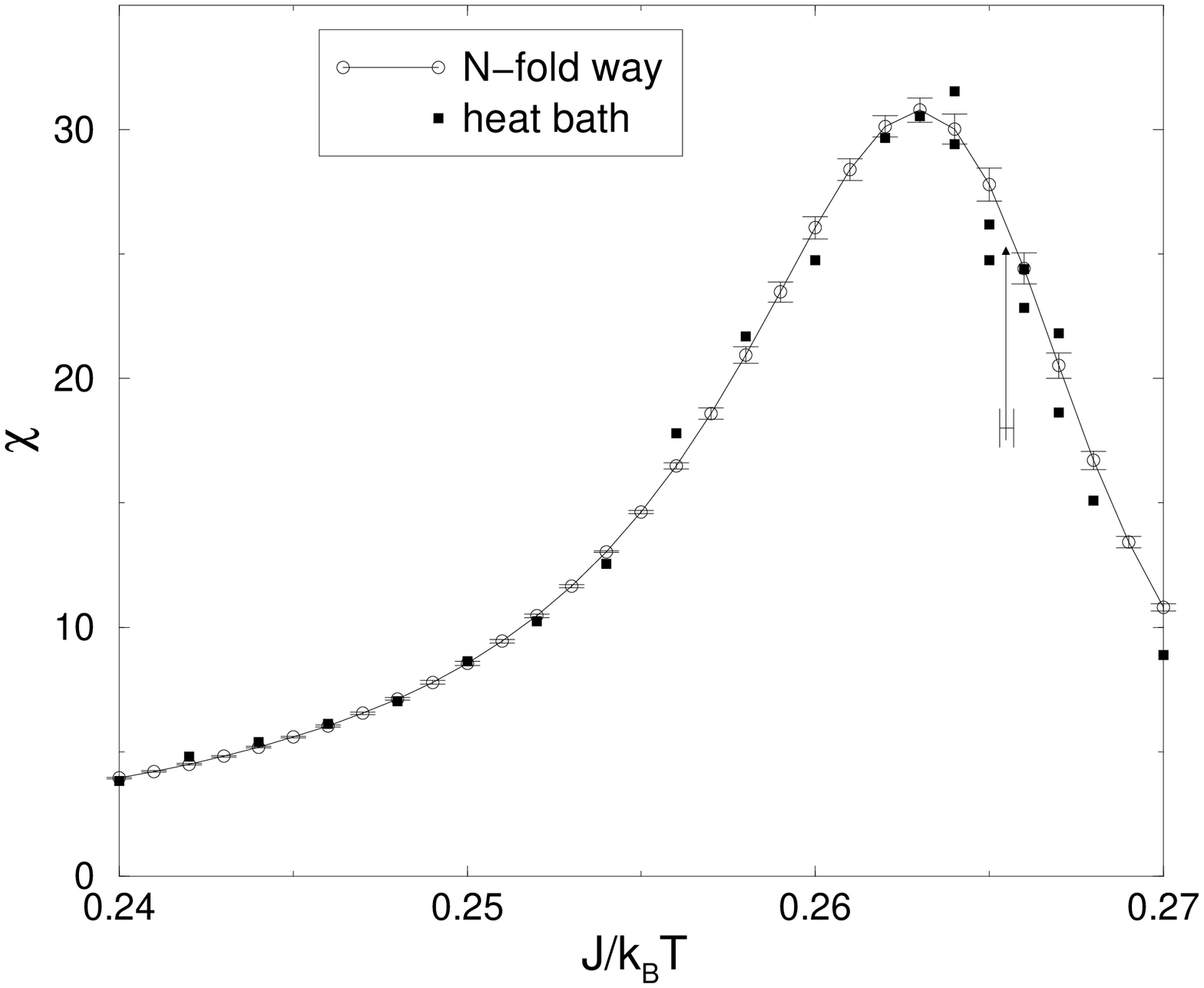}
\caption{Susceptibility $\chi$ of a thin Ising film with $L=32$ and $D=6$. The indicated error bars resulted from an average over $5$ runs. The average relative error of $\chi$ in the depicted range of $J/k_{B}T$ is $1.2\%$. The squares are data taken from ref. \cite{binderlandau}.}
\label{suscep}
\end{center}
\end{figure}

The two-dimensional Ising model is an ideal testing ground for Monte
Carlo algorithms since it can be solved exactly and therefore permits one to
verify simulational results directly. To show that our algorithm works also
well with other more complicated Ising systems, we have considered the following Hamiltonian

\begin{equation}
\mathcal{H}=-J\sum\limits_{\langle i,j\rangle}s_i s_j-H\!\!\!\sum\limits_{i\in
  \mbox{bulk}}\!\!\!s_i-h_1\!\!\!\!\!\!\!\!\sum\limits_{i\in
  \mbox{surface}\,1}\!\!\!\!\!\!\!\!s_i-h_2\!\!\!\!\!\!\!\!\sum\limits_{i\in
  \mbox{surface}\,2}\!\!\!\!\!\!\!\!s_i \quad \mbox{with}\quad s_i=\pm 1,
\label{thini}
\end{equation}
which is essentially a three-dimensional $L\times L\times D$ Ising model
confined by two walls with fields $h_1, h_2$ acting on them and periodic
boundary conditions in the $L\times L$ planes. A sketch of this geometry is depicted in figure \ref{thingeo}. 
With $H=0$ and $h_1=-h_2$ this model
exhibits the critical behavior of an interface localization/delocalization transition. For
finite $D$ one has a single transition at $T_{c}(D)$ which presumeably belongs
to the two-dimensional Ising universality class. For $T<T_{c}(D)$ the
interface is bound either to the left or the right wall and for
$T_{c}(D)<T<T_{cb}$ it is fluctuating delocalized around the center of the film, where
$T_{cb}$ is the critical bulk temperature. If $T$ satisfies $T>T_{cb}$ the film is disordered
with exception of the response to the surface fields $h_1, h_2$ near the walls.
It has become clear \cite{binderlandau,binderevans,ferrenberg} that such systems are extremely difficult to simulate, due to the fact that one has to deal with a much more severe slowing down than in simpler models, which results from the presence of a very large length scale $\xi_{\|}$ parallel to the surfaces in the phase with the delocalized interface which is related to the bulk correlation length $\xi_{b}$ via $\xi_{\|}\propto \exp (D/4\xi_{b})$ \cite{perryevans}. The corresponding correlation time is then described by $\tau \propto \xi_{\|}^{z}$, with a dynamic exponent of $z\approx 2$ \cite{hohenberg}. 
One can argue \cite{binderlandau} that the asymptotic region of the two-dimensional Ising critical behavior is very narrow and that the true nature of the transition can be observed only for $L\gg \xi_{\|}$. Consequently one is forced to study rather large linear dimensions $L$, depending on the thickness $D$ of the film, since for $D=12$, for example, one has to satisfy $L\gg \exp(D/4\xi_{b})=26$ ($\xi_{b}\approx 0.92$ for $J/k_{b}T=0.25$ \cite{liu,hasenbusch}). In the case considered here ($D=6$, $L=32$) we have $\exp(D/4\xi_{b})=5.1$ and the maximum of the specific heat is still rather broad, as can be seen from figure \ref{specH}. 
While for standard Ising models cluster-algorithms \cite{cluster1,cluster2} have proven to be a remedy for critical slowing down their utility has to be tested in detail when it comes to thin Ising films with additional surface fields. In \cite{dillmann} such a model has been studied with a ghostspin Swendsen-Wang cluster algorithm \cite{jswang3} and a significantly reduced auto-correlation time was achieved only for relatively small surface fields. Therefore we find it promising to
test the applicability of our new algorithm to the model in question. For this reason we have performed $5$ runs of a system with size $32\times 32\times6$ and competing surface fields $h_{1}/J=-h_{2}/J=0.55$. In order to calculate canonical averages of themodynamic observables we have estimated the density of states $\hat{g}(E)$
over the energy interval $\frac{E}{JN} \in\left[-2.8\bar{3},0.2\right]$ which consists of already $185974$ different energy levels despite the moderate size of the system. The whole interval was partitioned into $100$ adjacent intervals with small overlaps, which are 
needed to join the density of states afterwards. For the normalization of the
density of states we solely used the twofold degeneracy of the groundstate. 
The average total number of sweeps for the complete energy range stated above was $(1.14\pm 0.18)\cdot 10^{7}$, the average total CPU time amounted to $242.4\; h$ on a HP v-class, whereby a single energy interval needed $(8.7 \pm 0.2)\cdot 10^{3} s$ CPU time on average. The simulation was carried out in a parallel fashion such that the effective average running time was $12\; h$ for a complete run. The simulation was performed with $\epsilon=0.95$ and $f_{final}=(7.05\pm 0.03)\cdot 10^{-10}$.\\
In particular we have calculated the internal energy
\begin{equation}
U(T)=\langle E\rangle_{T}=\frac{\sum\limits_{E}E\hat{g}(E)e^{-\beta E}}{\sum\limits_{E}\hat{g}(E)e^{-\beta E}},
\end{equation}
the specific heat
\begin{equation}
C(T)=\frac{\langle E^2\rangle_{T}-\langle E\rangle_{T}^{2}}{T^2},
\end{equation}
as well as the Gibbs free energy
\begin{equation}
F(T)=-kT\ln \left(\sum\limits_{E}\hat{g}(E)e^{-\beta E} \right),
\end{equation}
and the entropy
\begin{equation}
S(T)=\frac{U(T)-F(T)}{T},
\end{equation}
over a range of temperatures near $T_{C}(D)$ and checked it against conventional Monte Carlo data obtained
from a heat bath algorithm \cite{binderlandau}.
We also considered canonical averages of the order parameter $\left| m
\right|=\left|\frac{1}{N}\sum s_i \right|$:
\begin{equation}
\langle \left|m \right| \rangle_{T}  = \frac{\sum\limits_{E}\langle \left|m\right| \rangle_{E}\, \hat{g}(E)e^{-\beta E}}{\sum\limits_{E}\hat{g}(E)e^{-\beta E}}, 
\end{equation}
its fourth order cumulant
\begin{equation}
U_{4}(T) =  1-\frac{\langle m^4 \rangle_{T}}{3\langle m^2 \rangle_{T}^{2}},
\end{equation}
and the susceptibility
\begin{equation}
\chi (T)  =  \frac{N}{T}\left(\langle m^2 \rangle_{T}- \langle \left|m\right| \rangle_{T}^{2}   \right),
\end{equation}
by sampling microcanonical averages $\langle \cdot \rangle_{E}$ during the last stage of the simulations. The results are shown in figure \ref{intE} - \ref{suscep}. The value of the extrapolated ($L\rightarrow \infty$) critical temperature $J/k_{B}T_{C}(D=6)=0.2655 \pm 0.0002$ estimated in ref. \cite{binderlandau} is marked by an arrow. One can see from these figures that quantities like the specific heat, the magnetization, the susceptibility and the fourth order cumulant can be obtained with an accuracy that is clearly better than that of the corresponding Monte Carlo data obtained with the standard single spin flip heat bath algorithm \cite{binderlandau}, although the effort in computer time was comparable. Of course, for a precise analysis of the critical behaviour of the present model it is necessary to repeat the study for several choices of $L$ and perform a finite size scaling analysis. This will be presented elsewhere \cite{schulz}. A further merit of the present approach is that it immediately yields precise data for the entropy and free energy of the model as well (figs. \ref{gibbsF},\ref{entropy}), which would be accessible from the standard algorithm \cite{binderlandau} only by tedious techniques of ``thermodynamic integration'' \cite{binder,frenkel,landau}. The availability of the free energy is particularly advantageous in the case of first-order interface localization-delocalization transitions \cite{ferrenberg,mueller}, where the relaxation time increases exponentially with the linear dimension $L$ of the system, and straightforward Monte Carlo studies would be hampered by hysteresis for large $L$ (and huge statistical errors for intermediate values of $L$ \cite{ferrenberg}). 

\section[con]{Conclusions} \label{sec5}
In this paper, it has been shown that the efficiency of the flat histogram method of Wang and Landau \cite{wanglandau} can be improved significantly by combining it with the so-called ``N-fold way'' \cite{bortz} technique of sampling states of a lattice model via Monte Carlo methods. In particular, the problem that errors get strongly enhanced near the right edge of an energy interval that is sampled is eliminated, and performing systematic comparisons for the two-dimensional Ising model with transition matrix Monte Carlo approaches proposed by Wang and Swendsen \cite{wangswendsen} it is shown that the present algorithm is competitive in efficiency with techniques described in the latter study \cite{wangswendsen}, but the merit of the present method is that it is rather straightforward to implement, and it should be immediately useful for wide classes of lattice models. Unlike cluster algorithms, there is no problem with the inclusion of magnetic fields. As a nontrivial example, we show new results of a first application to a thin Ising film with competing boundary fields.

\section*{Acknowledgement} 
One of us (B. J. S.) received financial support from the Deutsche Forschungsgemeinschaft (DFG) under grant No. Bi314/16-2. We are grateful to David P. Landau for numerous stimulating discussions, and for providing preprints of ref. \cite{wanglandau} prior to publication. We are grateful to Jian-Sheng Wang and Robert H. Swendsen for sending a preprint of ref. \cite{wangswendsen}. This work also received current support from the Bundesministerium f\"ur Bildung und Forschung (BMBF) under grant No. 03N6015.

\end{document}